# Predictive Control with Learning-Based Terminal Costs Using Approximate Value Iteration

**Francisco Moreno-Mora, Lukas Beckenbach, Stefan Streif** *

* *Technische Universität Chemnitz, Chemnitz, 09126 Germany.*
*Automatic Control and System Dynamics Lab (e-mail:*
*{francisco.moreno,lukas.beckenbach,stefan.streif}@etit.tu-chemnitz.de).*

**Abstract:** Stability under model predictive control (MPC) schemes is frequently ensured by terminal ingredients. Employing a (control) Lyapunov function as the terminal cost constitutes a common choice. Learning-based methods may be used to construct the terminal cost by relating it to, for instance, an infinite-horizon optimal control problem in which the optimal cost is a Lyapunov function. Value iteration, an approximate dynamic programming (ADP) approach, refers to one particular cost approximation technique. In this work, we merge the results of terminally unconstrained predictive control and approximate value iteration to draw benefits from both fields. A prediction horizon is derived in dependence on different factors such as approximation-related errors to render the closed-loop asymptotically stable further allowing a suboptimality estimate in comparison to an infinite horizon optimal cost. The result extends recent studies on predictive control with ADP-based terminal costs, not requiring a local initial stabilizing controller. We compare this controller in simulation with other terminal cost options to show that the proposed approach leads to a shorter minimal horizon in comparison to previous results.

*Keywords:* model predictive control; reinforcement learning; learning-based control.

## 1. INTRODUCTION

Model Predictive Control (MPC) is a control approach in which at each sampling time a cost function is minimized. The cost function involves the prediction of the system's trajectory, using the current system measurement as initial condition. An important characteristic of MPC is that constraint satisfaction is ensured by including them explicitly in the optimization problem.

Stability is an important property of MPC schemes. It is usually guaranteed by the inclusion of so-called terminal ingredients, a terminal cost, based on a control Lyapunov function and a terminal set, see Mayne et al. (2000) for a survey. The terminal cost may, for example, be found by a local linear quadratic regulation approach based on the linearized dynamics (Chen and Allgöwer, 1998) or constructed by a finite-horizon or infinite-horizon cost function under a known control law (see De Nicolao et al. (1998) and Magni et al. (2001); Köhler and Allgöwer (2021), respectively). Learning-based methods, which commonly involve a parametric approximation of an optimal infinite-horizon cost function, may also be used to design the terminal cost (Beckenbach and Streif, 2022).

Approximate dynamic programming (Lewis and Liu, 2013; Zhang et al., 2012) is one of such methods. It iteratively adapts an approximator to an infinite horizon optimal cost. Parametric approximators such as linear combinations of basis functions or neural networks are typically used for this task. The iteration schemes can roughly be classified in two categories, policy iteration (PI) and value iteration (VI) (see, e.g., Lewis and Liu (2013)). Policy iteration requires an initial stabilizing controller which is redundant in value iteration. Furthermore, finding this local control law may be intractable, which motivates the use of value iteration.

For either iteration scheme, cost convergence statements are numerous. Yet results such as Lincoln and Rantzer (2006); Bertsekas (2015); Heydari (2014) assume *exact* solution of the equations involved in the iteration scheme, which is rarely tractable in practice because parametric structures are used and equations are evaluated only at a finite number of sample points. Convergence properties and bounds of iterated costs accounting for approximation errors, which has been labeled as approximate value iteration, have been analyzed in, e.g., Bertsekas and Tsitsiklis (1996); Farahmand et al. (2010). In Heydari (2016), particular attention has been paid to the stability of the closed loop under the controller associated with the iterated approximate optimal cost which may be used in a predictive control context. A related approach has been presented recently in Beckenbach and Streif (2022) in which the aforementioned value iteration utilizes an initial controller. Based on the results of Köhler and Allgöwer (2021) and Heydari (2016), a minimal prediction horizon was derived, that guarantees asymptotic stability of the origin under the predictive controller in the presence of approximation errors.

Analogously, we propose a model predictive controller under an approximated terminal cost using value iteration. In contrast to Beckenbach and Streif (2022) and Köhler and Allgöwer (2021), we do not require an additional local exponentially stabilizing controller. By combining predictive control with value iteration, we combine the advantages of both approaches, namely direct constraint satisfaction and better suboptimality as measured by the infinite sum of the stage costs along the system trajectory. A bound of the suboptimality of the closed-loop system is also provided.

First, we define our problem setup, introduce approximate value iteration and present our model predictive control scheme in Section 2. Then, we present the stability analysis under the proposed controller in Section 3. In Section 4 we analyze the behavior of the closed-loop system in simulation and compare the proposed controller with other approaches. Section 5 concludes the paper.

*Notation.* We denote the set of integers greater or equal to an integer $i$ by $\mathbb{Z}_i$ and the set of non-negative real numbers by $\mathbb{R}_{\geq 0}$. The ceiling function is denoted by $\lceil \cdot \rceil$. $\mathcal{K}_\infty$ denotes the set of functions $\alpha : \mathbb{R}_{\geq 0} \to \mathbb{R}_{\geq 0}$ that are continuous, strictly increasing, unbounded and satisfy $\alpha(0) = 0$. The Kronecker product is denoted by $\otimes$ and $\mathrm{diag}(x_1, \ldots, x_n)$ is a diagonal matrix with main diagonal entries $x_1, \ldots, x_n$.

## 2. PROBLEM SETUP AND PRELIMINARIES

We consider a discrete-time nonlinear system
$$x(k+1) = f(x(k), u(k)), \quad k \in \mathbb{Z}_0 \quad (1)$$
where $x(k) \in \mathbb{R}^n$ is the state, with initial value $x(0) = x_0$. $u(k) \in \mathbb{R}^m$ is the control input and $f$, satisfying $f(0,0) = 0$, is continuous. The system is subject to constraints
$$(x(k), u(k)) \in \mathbb{D} := \mathbb{X} \times \mathbb{U} \subset \mathbb{R}^n \times \mathbb{R}^m, \quad k \in \mathbb{Z}_0,$$
for a compact $\mathbb{D}$, with $0 \in \mathrm{int}(\mathbb{D})$. In this work, predictive control is associated with a continuous stage cost $l : \mathbb{R}^n \times \mathbb{R}^m \to \mathbb{R}_{\geq 0}$ of the form
$$l(x, u) = Q(x) + u^\top R u, \quad Q : \mathbb{R}^n \to \mathbb{R}_{\geq 0}, \quad R \succ 0.$$
The following assumption is required, which can be fulfilled, e.g., through quadratic stage costs.

*Assumption 1.* There exist functions $\underline{\alpha}_l, \overline{\alpha}_l \in \mathcal{K}_\infty$, such that $\underline{\alpha}_l(\|x\|) \leq l(x, 0) \leq \overline{\alpha}_l(\|x\|)$ for all $x \in \mathbb{R}^n$.

Prior to introducing the predictive controller, value iteration and its approximate version are discussed next.

### 2.1 Value Iteration

Value iteration is a function approximation approach for unconstrained optimal control problems. In the current context, the subsequent procedure may thus only be valid in a sufficiently small neighborhood of the origin. Define the infinite-horizon cost for some $x \in \mathbb{R}^n$
$$J(x, \overline{u}(\cdot)) := \sum_{k=0}^{\infty} l(\overline{x}(k), \overline{u}(k))$$
with $\overline{x}(k+1) = f(\overline{x}(k), \overline{u}(k))$, $\overline{x}(0) = x$ and define the value function as its minimum (tacitly assuming its existence), i.e.,
$$V(x) := \min_{\overline{u}(\cdot)} J(x, \overline{u}(\cdot)).$$

*Definition 1.* (Heydari (2016)) An admissible controller within a set $\Omega \subset \mathbb{R}^n$, containing the origin, is defined as a continuous law $u : \Omega \to \mathbb{R}^m$ with $u(0) = 0$ such that $J(x, \overline{u}(\cdot))$, with $\overline{u}(k) = u(x(k))$, $k \in \mathbb{Z}_0$, is finite for any $x \in \Omega$.

To ensure boundedness of the value function on $\Omega$, existence of at least one admissible controller is required. Take $\Omega \subset \mathbb{R}^n$ compact with $0 \in \mathrm{int}(\Omega)$.

*Assumption 2.* There exists an admissible controller for the system (1) on $\Omega$.

The value function satisfies the Bellman equation
$$V(x) = \min_u l(x, u) + V(f(x, u))$$
with associated optimal controller given by
$$h^*(x) = \arg\min_u l(x, u) + V(f(x, u)).$$

Value iteration describes an iterative procedure to approximate the value function over a domain of interest, in our case $\Omega \subset \mathbb{R}^n$. It is done by performing
$$V^{i+1}(x) = \min_u l(x, u) + V^i(f(x, u)), \quad \forall x \in \Omega \quad (2)$$
on the auxiliary approximator $V^i : \Omega \to \mathbb{R}_{\geq 0}$ over $i = 0, 1, \ldots$, with an initial guess $V^0$. Convergence of value iteration has been shown for different conditions, see e.g. Lincoln and Rantzer (2006); Heydari (2014).

### 2.2 Approximate Value Iteration

Performing the iteration defined by (2) amounts to solving the equation *exactly* over $\Omega$ which is practically tractable only in special cases such as, e.g., a linear quadratic regulator problem.

Define the approximation error at the $i$-th iteration as $e^i : \Omega \to \mathbb{R}$. Henceforth, we relax (2) and consider iterations of the form
$$\hat{V}^{i+1}(x) = \min_u l(x, u) + \hat{V}^i(f(x, u)) + e^i(x), \\ \forall x \in \Omega. \quad (3)$$

We associate a controller with each $\hat{V}^i$ as follows
$$h^i(x) := \arg\min_u l(x, u) + \hat{V}^i(f(x, u)), \quad \forall x \in \Omega. \quad (4)$$

Commonly (Granzotto et al., 2021), value iteration is performed until a convergence tolerance $\delta(x) > 0$ has been reached, frequently specified in the form
$$|\hat{V}^{i+1}(x) - \hat{V}^i(x)| \leq \delta(x), \quad \forall x \in \Omega. \quad (5)$$

It is worth highlighting that neither $e^i$ nor $\delta$ represent the error between $V^i$ and the approximate $\hat{V}^i$ or the error between the iterated cost to the (in general unknown) value function $V$.

*Remark 1.* The convergence criterion (5) may have a specific form that allows extracting further information about the relationship between $V^i$ and $V$ (in the ideal case), see e. g. Granzotto et al. (2021).

The following result, which contains a minor modification of (Heydari, 2016, Theorem 4), shows asymptotic stability of the system under the controller $\hat{h}^i$ when the approximation error and the tolerance are sufficiently small.

*Lemma 2.* Let $\hat{V}^i : \Omega \to \mathbb{R}_{\geq 0}$ be continuously differentiable and the iteration error be bounded by $|e^i(x)| \leq c_e l(x,0)$, for some $c_e > 0$ and $\forall i \in \mathbb{Z}_0$. Suppose there exist $c_\delta > 0$ such that $|\delta(x)| \leq c_\delta l(x,0)$, $\forall x \in \Omega$. If $c_e + c_\delta < 1$, then the controller $h^i(x)$ resulting from the approximate value iteration terminated with tolerance $\delta(x)$ asymptotically stabilizes the system for any initial state in $\mathcal{B}_{\overline{r}}$, where $\mathcal{B}_r := \{x \in \mathbb{R}^n : \hat{V}^i(x) \leq r\}$ and $\overline{r}$ is the largest $r$ such that $\mathcal{B}_r \subset \Omega$. Specifically, it holds that

$$\hat{V}^i(f(x,h^i(x))) - \hat{V}^i(x) \leq (c_e + c_\delta - 1)l(x,0), \quad \forall x \in \Omega. \quad (6)$$

**Proof.** We proceed in a similar manner as done in the proof of (Heydari, 2016, Theorem 4), where $\hat{V}^i$ is used as a Lyapunov function candidate. Using (3) and (4) in (5) we obtain

$$\hat{V}^i(f(x,h^i(x))) - \hat{V}^i(x) \leq -l(x,h^i(x)) - e^i(x) + \delta(x), \\ \forall x \in \Omega.$$

It follows immediately that

$$\hat{V}^i(f(x,h^i(u))) - \hat{V}^i(x) \leq (c_e + c_\delta - 1)l(x,0), \quad \forall x \in \Omega.$$

By assumption, $c_e + c_\delta < 1$ and thus asymptotic stability follows for any initial state in the largest sub-level set of $\hat{V}^i$ in $\Omega$.

Hence, a local controller is derived for the nonlinear system. However, fulfilling the constraints and the required approximation error and convergence tolerance to obtain a stabilizing controller, which requires a suitable selection of the approximator $\hat{V}^i$, may be only possible for an undesirably small $\Omega$. Yet, constraint satisfaction and enlargement of the region of attraction can be achieved by embedding the resulting cost in a predictive control framework, as demonstrated in the subsequent section.

### 2.3 VI-based Model Predictive Control

The herein inspected predictive controller is based on the following optimization problem

$$V_N(x) := \min_{\overline{u}(\cdot;x)} \sum_{j=0}^{N-1} l(\overline{x}(j;x),\overline{u}(j;x)) + \hat{V}^i(\overline{x}(N;x)) \quad (7a)$$

$$\text{s.t. } \overline{x}(j+1;x) = f(\overline{x}(j;x),\overline{u}(j;x)), \ \overline{x}(0;x) = x \quad (7b)$$

$$(\overline{x}(j),\overline{u}(j)) \in \mathbb{D}, \quad \forall j \in \{0,\ldots,N-1\}, \quad (7c)$$

where $\hat{V}^i$ is the result of the approximate value iteration (3) for some convergence tolerance $\delta$ as specified in (5) satisfying $|\delta(x)| \leq c_\delta l(x,0)$ for some $c_\delta > 0$. Denote the solution of the problem as $\overline{u}^*(\cdot;x)$ with associated state trajectory $\overline{x}^*(\cdot;x)$. At each time step index, solve (7) at the current state of the system $x = x(k)$ and apply the controller $u(k) = \overline{u}(0;x(k))$ to the system.

The next section contains the stability results. Instead of enforcing a terminal condition, as frequently done, a horizon length $N \in \mathbb{N}$ is chosen that allows us to exploit the learned value function $\hat{V}^i$ to guarantee asymptotic stability of the origin under $u(k)$, constraint satisfaction and recursive feasibility.

## 3. MAIN RESULTS

Similar to Grüne (2012); Boccia et al. (2014), a local bound on the cost function is assumed as follows.

*Assumption 3.* There exists a constant $\gamma > 0$ such that $V_N(x) \leq \gamma l(x,0), \forall x \in \Omega$.

The subsequent property allows us to exploit the analysis of Köhler and Allgöwer (2021) to determine a sufficiently long prediction horizon.

*Assumption 4.* There exists a constant $\varepsilon > 0$ such that the set $\mathcal{B}_\varepsilon := \{x \in \mathbb{R}^n \,|\, l(x,0) \leq \varepsilon\}$ is contained in $\Omega$ and $(x,h^i(x)) \in \mathbb{D}$, $\forall x \in \mathcal{B}_\varepsilon$.

*Remark 3.* It can be shown in a similar way as in (Heydari, 2016, Theorem 1) that if the value iteration is performed using continuous approximators and the initial guess $V^0$ is continuous, then,

$$V_N^i(x) := \sum_{k=0}^{N} l(\hat{x}(k),h^i(\hat{x}(k))) + \hat{V}^i(\hat{x}(N)),$$

with $\hat{x}(k+1) := f(\hat{x}(k),h^i(\hat{x}(k)))$, $\hat{x}(k) = x$, is continuous over $\Omega_\beta := \{x \in \Omega | \hat{V}^i(x) \leq \beta\}$, for $\beta > 0$ such that $\Omega_\beta \subseteq \Omega$. The set $\Omega_\beta$ includes the origin. Furthermore $V_N^i(x)$ is bounded in $\Omega \setminus \Omega_\beta$. Thus, in this case an upper bound as required by the assumption can be determined. Regarding Assumption 4, for a nonempty $\Omega_\beta$ and $l$ satisfying Assumption 1, an $\varepsilon > 0$ can be found.

Since the properties of $\hat{V}^i$ are valid only on $\Omega$, the terminal predicted state is required to lie in this set. The set $\mathcal{B}_\varepsilon$ can be used to enforce this.

*Lemma 4.* (Terminal state) Let Assumptions 1–4 hold. Then, for any $\overline{V} > 0$ there exists $N_\Omega \in \mathbb{N}$ such that for all $N \in \mathbb{N}$, $N \geq N_\Omega$ and any $x \in \mathbb{X}_{\overline{V}} := \{y \in \mathbb{X} | V_N(y) \leq \overline{V}\}$, it holds that $\overline{x}^*(N;x) \in \mathcal{B}_\varepsilon$. Additionally, $V(N;x) \leq \rho_\gamma^{N-N_0} \min\{\gamma l(0,\overline{u}^*(0;x)),\underline{\gamma}\varepsilon\}$ holds, where $N_0 = \left\lceil \max\left\{0, \frac{\overline{V}-\gamma\varepsilon}{\varepsilon}\right\}\right\rceil$ and $V(k;x) = V_{N-k}(\overline{x}^*(k;x))$, with $\rho_\gamma = \frac{\gamma-1}{\gamma}$ and $\underline{\gamma} = \min\{\gamma,\overline{V}/\varepsilon\}$.

A proof of this lemma can be found in (Köhler and Allgöwer, 2021, Theorem 5), as the same results are valid, but using $\gamma$ as defined in Assumption 3 and $V_N$ as defined in (7), because $l_{\min}(x) = l(x,0)$, with $l_{\min}(x) := \inf_{u \in \mathbb{U}} l(x,u)$. As seen therein,

$$N_\Omega = N_0 + \left\lceil \frac{\max\{\log(\underline{\gamma}),0\}}{\log(\gamma) - \log(\gamma - 1)} \right\rceil \quad (8)$$

yields the desired properties.

Not only must $N$ be chosen in such a way, that the terminal state is in the required set, but it also must be large enough to guarantee asymptotic stability of the closed loop, as specified by the next theorem.

*Theorem 5.* Let Assumptions 1–4 hold. Suppose $|e^j(x)| \leq c_e l(x,0)$, $\forall j \in \{0,\ldots,i\}$, and the convergence tolerance is bounded by $|\delta(x)| \leq c_\delta l(x,0)$, with $c_{e,\delta} > 0$ such that $c_e + c_\delta < 1$. Consider $N_\Omega$ as defined in (8). Then, there exists $N_{\overline{V}} \in \mathbb{N}$ satisfying $N_{\overline{V}} \geq N_\Omega$ such that for any $x_0 \in \mathbb{X}_{\overline{V}} := \{y \in \mathbb{R}^n \,|\, V_N(y) \leq \overline{V}\}$ the predictive control problem (7) is feasible for all $k \in \mathbb{Z}_0$, the constraints are satisfied, and the origin is asymptotically stable for the resulting closed loop. Furthermore, there exists a constant $\alpha \in (0,1]$ such that the infinite-horizon performance is bounded as follows

$$\sum_{k=0}^{\infty} l(x(k), u(k)) \leq \frac{1}{\alpha} V_\infty(x_0), \quad (9)$$

with

$$V_\infty(x) := \min_{\overline{u}(\cdot;x)} \sum_{j=0}^{\infty} l(\overline{x}(j;x), \overline{u}(j;x)) \quad (10a)$$

$$\text{s.t. } \overline{x}(j+1;x) = f(\overline{x}(j;x), \overline{u}(j;x)), \ \overline{x}(0;x) = x \quad (10b)$$

$$(\overline{x}(j), \overline{u}(j)) \in \mathbb{D}, \quad \forall j \in \mathbb{Z}_0. \quad (10c)$$

**Proof.** The proof is similar to standard arguments for MPC with local Lyapunov functions as terminal cost. First, notice that the following inequalities hold

$$l(x,0) \leq V_N(x) \leq \gamma_{\overline{V}} l(x,0), \quad (11)$$

where $\gamma_{\overline{V}} := \max\{\gamma, \overline{V}/\varepsilon\}$, for $x \in \mathbb{X}_{\overline{V}}$. The lower bound follows from the properties of the stage cost and the upper bound can be obtained by using a case distinction as in Köhler and Allgöwer (2021), whether $x \in \mathcal{B}_\varepsilon$ or not. Consider the difference $\Delta V_N := V_N(x(k+1)) - V_N(x(k))$, then

$$\Delta V_N = \sum_{j=0}^{N-1} l(\overline{x}^*(j;x(k+1)), \overline{u}^*(j;x(k+1)))$$
$$- l(\overline{x}^*(j;x(k)), \overline{u}^*(j;x(k)))$$
$$+ \hat{V}^i(\overline{x}^*(N;x(k+1))) - \hat{V}^i(\overline{x}^*(N;x(k))).$$

Consider an input sequence $u(\cdot; x(k+1))$ such that $u(j;x(k+1)) = \overline{u}^*(j+1;x(k))$ for $j \in \{0,\ldots,N-2\}$ with associated system trajectory $x(j;x(k+1))$. Then, $u(j;x(k+1))$ is a feasible input sequence, thus

$$\Delta V_N \leq -l(x(k),u(k)) + l(x^*(N;x(k)), u(N-1;x(k+1)))$$
$$+ \hat{V}^i(x(N;x(k+1))) - \hat{V}^i(x^*(N;x(k))).$$

Assuming $N \geq N_\Omega$ from Lemma 4, it holds that $\overline{x}^*(N;x(k)) \in \mathcal{B}_\varepsilon$. Then, with $u(N-1;x(k+1)) = h^i(x(N-1;x(k+1))) = h^i(x^*(N;x(k)))$ and (6) we obtain

$$\Delta V_N \leq -l(x(k),u(k)) + (c_e + c_\delta) l(x^*(N;x(k)),0)$$
$$\overset{(6)}{\leq} -l(x(k),u(k)) + \underbrace{\frac{c_e + c_\delta}{1 - (c_e + c_\delta)}}_{=:c_N} V(N|t)$$
$$\leq -l(x(k),u(k)) + c_N \rho_\gamma^{N-N_0} \gamma l(x(k),u(k)).$$

Hence, for

$$N > \underline{N} := N_0 + \left\lceil \frac{\log(c_N) + \log(\gamma)}{\log(\gamma) - \log(\gamma-1)} \right\rceil,$$

$V_N(x(k))$ decreases along the system trajectory and $x(k) \in \mathbb{X}_{\overline{V}}$, for $k \in \mathbb{Z}_0$. Considering $N_\Omega$ and $\underline{N}$, selecting

$$N > N_{\overline{V}} := N_0 + \left\lceil \frac{\max\{\log(\gamma_{\overline{V}}), \log(c_N) + \log(\gamma), 0\}}{\log(\gamma) - \log(\gamma-1)} \right\rceil$$

guarantees asymptotic stability of the origin.

Furthermore, it is known from Heydari (2016) that $\hat{V}^i(x) \leq \overline{V}^\infty(x)$ with

$$\overline{V}^\infty(x) = \min_{\hat{u}(\cdot;x)} \sum_{k=0}^{\infty} l(\hat{x}(k;x), \hat{u}(k;x)) + c_e l(\hat{x}(k;x),0)$$

and $\hat{x}(k+1;x) = f(\hat{x}(k;x), \hat{u}(k;x)), \ \hat{x}(0;x) = x$.

Using $l(x,0) \leq l(x,u), \forall u \in \mathbb{R}^m$, we obtain the following bound

$$V_N(x)$$
$$\leq \min_{\substack{\hat{u}(\cdot;x) \\ \text{s.t.}(7b),(7c)}} \sum_{j=0}^{N-1} l(\hat{x}(j;x), \hat{u}(j;x)) + \overline{V}^\infty(\hat{x}(N;x))$$
$$\leq (1+c_e) \Bigg( \min_{\substack{\hat{u}(\cdot;x) \\ \text{s.t.}(7b),(7c)}} \sum_{j=0}^{N-1} l(\hat{x}(j;x), \hat{u}(j;x))$$
$$+ \min_{\substack{\hat{u}(\cdot;\hat{x}(N;x)) \\ \text{s.t.}(10b),(10c)}} \sum_{m=0}^{\infty} l(\hat{x}(m;\hat{x}(N;x)), \hat{u}(m;\hat{x}(N;x))) \Bigg)$$
$$= (1+c_e) V_\infty(x).$$

Summing $\Delta V$ from 0 to infinity we obtain

$$c_V \sum_{j=0}^{\infty} l(x(k),u(k)) \leq V_N(x_0),$$

where $c_V := 1 - c_N \rho_\gamma^{N-N_0} \gamma$. Inequality (9) holds with $\alpha := \frac{1+c_e}{c_V}$.

*Remark 6.* A key point of our approach is the set $\mathcal{B}_\varepsilon$, where the control policy associated with the approximated cost *fulfills the constraints*. For sufficiently small sets and locally stabilizable linearized dynamics, the local approximation is close to a local linear quadratic regulator solution as utilized in, e.g., Chen and Allgöwer (1998) which may eliminate the need for a nonlinear analysis.

## 4. NUMERICAL STUDY

In this section we apply the theoretical results to an orbital maneuver problem which has been addressed using ADP in Heydari and Balakrishnan (2014) and Beckenbach and Streif (2022). The state of the system is selected as $x(k) = [X(k), Y(k), X_t(k), Y_t(k)]^\top$, where $[X,Y]^\top$ describes the position of the spacecraft measured from the orbital frame in the destination orbit and $[X_t, Y_t]^\top$ its velocity. The discretized dynamics of the system are given by

$$x(k+1) = x(k)$$
$$+ \Delta t \begin{bmatrix} x_3(k) \\ x_4(k) \\ 2x_4(k) - (1+x_1(k))\left(\frac{1}{r(k)^3} - 1\right) \\ -2x_3(k) - x_2(k)\left(\frac{1}{r(k)^3} - 1\right) \end{bmatrix} + \Delta t \begin{bmatrix} 0 & 0 \\ 0 & 0 \\ 1 & 0 \\ 0 & 1 \end{bmatrix} u(k)$$

with $r(k) = \sqrt{(1+x_1(k))^2 + x_2(k)^2}$. We consider constraints $\mathbb{X} = [-0.5, 0.5]^4$, $\mathbb{U} = [-2,2]^2$ and set the sampling time to $\Delta t = 0.05$. For the weighting matrices we select $Q = \text{diag}(50,50,50,50)$ and $R = \text{diag}(1,1)$, which result in relatively small $\gamma$ and large $\varepsilon$. For the approximate value iteration we consider the domain $\Omega = [-0.12 \ 0.12]^4$ and select 80 samples from a uniform distribution over the domain. The approximation error and convergence tolerance are evaluated using $10^3$ samples, selected also from a uniform distribution. We select the approximator as a linear combination of basis functions as in Heydari (2016) and Beckenbach and Streif (2022)

$$\hat{V}^i(x) = w_i^T [(x \otimes x)^\top \ (x \otimes x \otimes x)^\top]^\top.$$

Thus, the basis functions are comprised of all the monomials up to third degree of the state variables.

First we compare in Fig. 1 the resulting stabilizing horizon for different values of the approximation error and the convergence tolerance, measured by $c_e$ and $c_\delta$ respectively for $x_0 \in \mathbb{X}$. It can be seen that for approximately $c_e + c_\delta \leq 0.36$ the minimum horizon remains at 127 and that for values greater than approximately 0.64 it rapidly increases to 210 for a value of 0.97. It is worth mentioning that we look at the *combined* influence of the approximation error and the convergence tolerance, as solving (3) with a very small error is relatively uncomplicated. Using, the *fminunc* solver from Matlab required just one or two solver iterations.

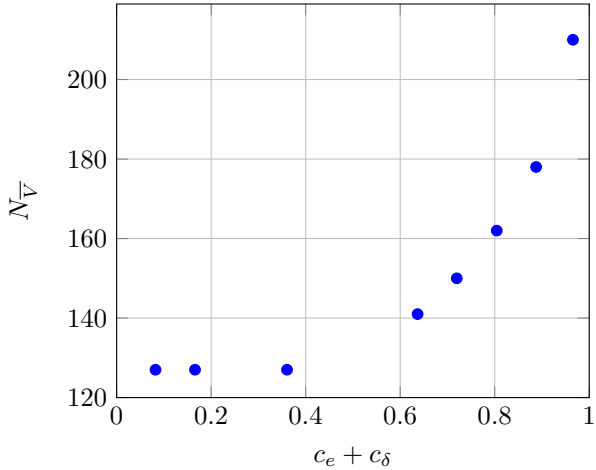

Fig. 1. Stabilizing horizon $N_{\overline{V}}$ for different values of $c_e + c_\delta$.

Applying the results in Köhler and Allgöwer (2021) using a LQR controller for the linearized system as the local stabilizing controller results in $C = 3.66$, $\rho = 0.91$ and $\varepsilon = 0.5$, where the value of $\varepsilon$ is the largest for which the LQR controller fulfills the input constraints. This results in a minimum stabilizing horizon of 210. Even though the shortest prediction horizon derived from our results is almost 40% shorter than the one obtained using the LQR controller and significantly shorter than the results in Beckenbach and Streif (2022), the required horizon is still conservative. This is seen in Fig. 2, where the resulting trajectory and control input when applying the proposed MPC controller were plotted using a horizon of $N = 40$ for different values of the approximation error and convergence tolerance.

It can be seen that there is almost no difference in the resulting trajectories and control inputs. This is also the case for $N > 40$, as the predicted terminal state is very close to the origin for longer horizons.

Even though the minimum stabilizing horizon derived is conservative, the benefits of including a learned terminal cost can be seen in Fig. 3. Here, the states and the resulting input for the proposed MPC controller and an MPC controller *without* terminal costs were plotted; in both cases for $N = 10$. It can be seen that the ADP-based controller steers the system to the origin in 100 simulation time steps, whereas the system under the MPC controller without terminal costs is still far away from the origin. A horizon of $N = 45$ would be necessary to produce, almost, the same results obtained with our proposed ADP-based MPC controller, but with the MPC controller without terminal cost.

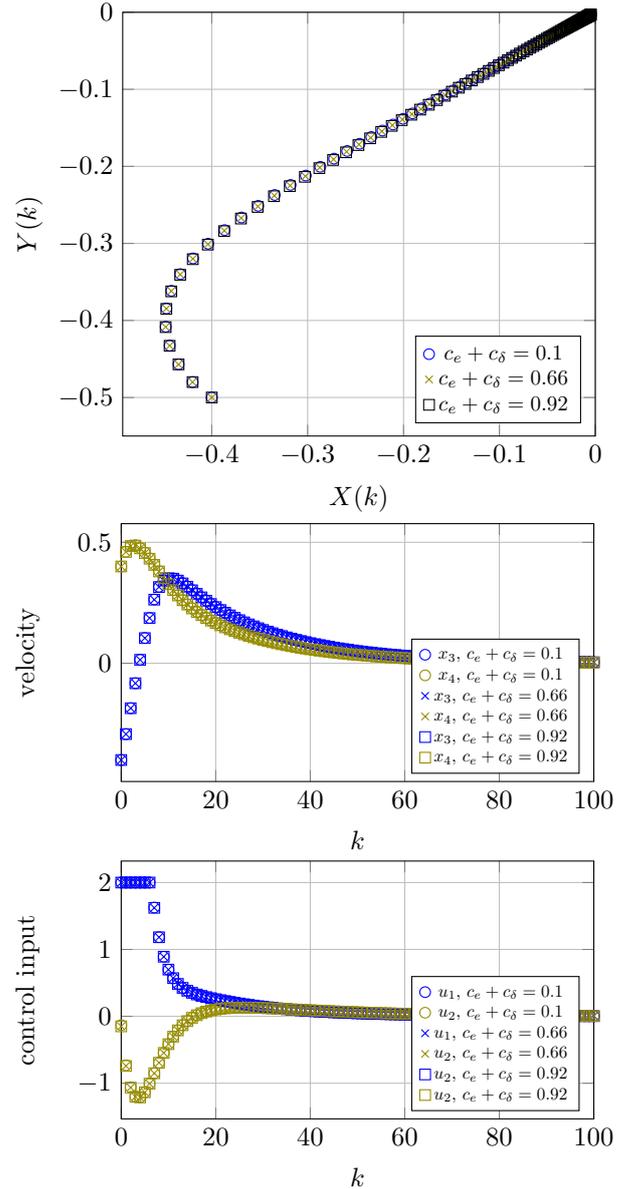

Fig. 2. Position, speed and control input of the resulting closed-loop system for different values of $c_e + c_\delta$.

## 5. CONCLUSION

We presented a model predictive controller that employs the approximation of the value function of an infinite horizon optimal control problem, obtained by using approximate value iteration, as a terminal cost. This approach allows us to consider approximation errors to analyze the properties of the learned costs and its associated controller. We then determine a minimal prediction horizon to exploit the approximated value function without the need of a terminal constraint. The horizon length is affected by the user-defined convergence tolerance and the approximation error as well the domain of approximation and the domain of interest. Thus, the proposed approach is suitable for the case where the control law associated with the approximated value function leads to a feasible system trajectory for a large subset of the feasible set, otherwise our approach reduces to using a quadratic approximation of the value function using a LQR controller as a locally stabilizing controller. Even though the required prediction horizon can be reduced in comparison to other related approaches, a certain conservatism can be observed which is to be improved in future works.

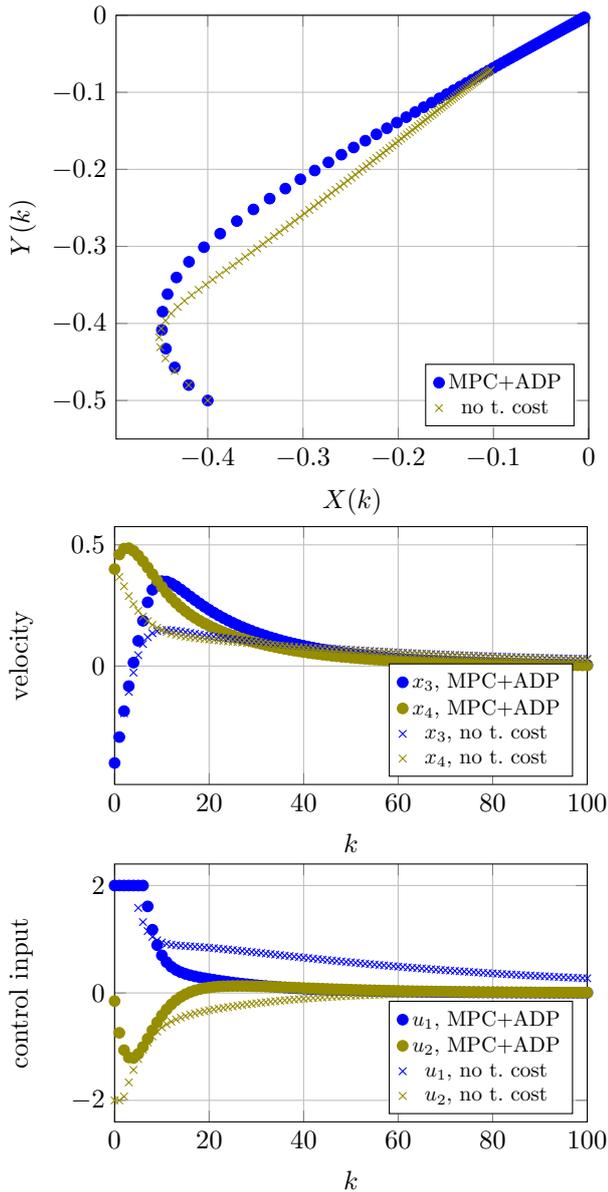

Fig. 3. Position, speed and control input of the resulting closed-loop system for the proposed ADP-based MPC controller and an MPC controller without terminal costs.

## REFERENCES

Beckenbach, L. and Streif, S. (2022). Approximate infinite-horizon predictive control. In *Proc. of the 61st IEEE Conference on Decision and Control*. To appear, available at arXiv:2111.08319.

Bertsekas, D. and Tsitsiklis, J.N. (1996). *Neuro-dynamic programming*. Athena Scientific.

Bertsekas, D.P. (2015). Value and policy iterations in optimal control and adaptive dynamic programming. *IEEE transactions on neural networks and learning systems*, 28(3), 500–509.

Boccia, A., Grüne, L., and Worthmann, K. (2014). Stability and feasibility of state constrained MPC without stabilizing terminal constraints. *Systems & Control Letters*, 72, 14–21.

Chen, H. and Allgöwer, F. (1998). A quasi-infinite horizon nonlinear model predictive control scheme with guaranteed stability. *Automatica*, 34(10), 1205–1217.

De Nicolao, G., Magni, L., and Scattolini, R. (1998). Stabilizing receding-horizon control of nonlinear time-varying systems. *IEEE Transactions on Automatic Control*, 43(7), 1030–1036.

Farahmand, A.m., Szepesvári, C., and Munos, R. (2010). Error propagation for approximate policy and value iteration. In *Advances in Neural Information Processing Systems*, volume 23. Curran Associates, Inc.

Granzotto, M., Postoyan, R., Nešić, D., Buşoniu, L., and Daafouz, J. (2021). When to stop value iteration: stability and near-optimality versus computation. In *Proc. of Machine Learning Research*, volume 144, 412–424.

Grüne, L. (2012). NMPC without terminal constraints. *IFAC Proceedings Volumes*, 45(17), 1–13.

Heydari, A. (2014). Revisiting approximate dynamic programming and its convergence. *IEEE transactions on cybernetics*, 44(12), 2733–2743.

Heydari, A. (2016). Theoretical and numerical analysis of approximate dynamic programming with approximation errors. *Journal of Guidance, Control, and Dynamics*, 39(2), 301–311.

Heydari, A. and Balakrishnan, S. (2014). Adaptive critic-based solution to an orbital rendezvous problem. *Journal of Guidance, Control, and Dynamics*, 37(1), 344–350.

Köhler, J. and Allgöwer, F. (2021). Stability and performance in mpc using a finite-tail cost. *IFAC-PapersOnLine*, 54(6), 166–171.

Lewis, F.L. and Liu, D. (2013). *Reinforcement learning and approximate dynamic programming for feedback control*. John Wiley & Sons.

Lincoln, B. and Rantzer, A. (2006). Relaxing dynamic programming. *IEEE Transactions on Automatic Control*, 51(8), 1249–1260.

Magni, L., De Nicolao, G., Magnani, L., and Scattolini, R. (2001). A stabilizing model-based predictive control algorithm for nonlinear systems. *Automatica*, 37(9), 1351–1362.

Mayne, D.Q., Rawlings, J.B., Rao, C.V., and Scokaert, P.O. (2000). Constrained model predictive control: Stability and optimality. *Automatica*, 36(6), 789–814.

Zhang, H., Liu, D., Luo, Y., and Wang, D. (2012). *Adaptive dynamic programming for control: algorithms and stability*. Springer Science & Business Media.